# Observation of Fermi-surface-dependent anisotropic Cooper pairing in kagome superconductor $CsV_3Sb_5$

Akifumi Mine[1,†], Yigui Zhong[1,†,*], Jinjin Liu[2,3,†], Takeshi Suzuki[1], Sahand Najafzadeh[1], Takumi Uchiyama[1], Jia-Xin Yin[4], Xianxin Wu[5], Xun Shi[2,3], Zhiwei Wang[2,3,6,*], Yugui Yao[2,3,6], and Kozo Okazaki[1,7,8,*]

[1]*Institute for Solid State Physics, The University of Tokyo, Kashiwa, Japan*

[2]*Centre for Quantum Physics, Key Laboratory of Advanced Optoelectronic Quantum Architecture and Measurement (MOE), School of Physics, Beijing Institute of Technology, Beijing, China*

[3]*Beijing Key Lab of Nanophotonics and Ultrafine Optoelectronic Systems, Beijing Institute of Technology, Beijing, China*

[4]*Department of Physics, Southern University of Science and Technology, Shenzhen, China*

[5]*CAS Key laboratory of Theoretical Physics, Institute of Theoretical Physics, Chinese Academy of Sciences, Beijing, China*

[6]*Material Science Center, Yangtze Delta Region Academy of Beijing Institute of Technology, Jiaxing, China*

[7]*Trans-scale Quantum Science Institute, The University of Tokyo, Tokyo, Japan*

[8]*Material Innovation Research Center, The University of Tokyo, Kashiwa, Japan*

[†]Equally contributed authors.
*Corresponding authors: yigui-zhong@issp.u-tokyo.ac.jp; zhiweiwang@bit.edu.cn; okazaki@issp.u-tokyo.ac.jp

## Abstract

In the recently discovered kagome superconductor $AV_3Sb_5$ ($A$ = K, Rb, and Cs), superconductivity is intertwined with an unconventional charge density wave order. The pairing symmetry remains elusive owing to the lack of direct measurement of the superconducting gap in the momentum space. Here, utilizing laser-based ultra-high-resolution and low-temperature angle-resolved photoemission spectroscopy, we observe Fermi-surface-dependent anisotropic Cooper pairing in kagome superconductor $CsV_3Sb_5$. We detect a highly anisotropic superconducting gap structure with anisotropy exceeding 80% and a gap maximum along the V-V bond direction on a Fermi surface originating from the 3*d*-orbital electrons of the V kagome lattice. This is in stark contrast to the isotropic superconducting gap structure on the Fermi surface occupied by Sb 5*p*-orbital electrons. Our direct observation of the Fermi-surface-dependent anisotropic pairing in $CsV_3Sb_5$ is fundamental for understanding the intertwined orders in the ground state of kagome superconductors.

## Introduction

The unique geometry of the kagome lattice, consisting of corner-shared triangles and hexagons, leads to the geometrical frustration in the spin degree of freedom and the destructive quantum interference in the electronic wavefunctions [1-3]. The latter generates the localized density of states inside the hexagons, which introduces a topological flat band. Additionally, because of the similar group symmetry with graphene, the kagome lattice materials exhibit a Dirac cone and van Hove singularities (VHS) in their electronic band structure. Thereby, kagome lattice materials provide an ideal platform for studying the interplay between topology, correlations, and emergent electronic orders [4-6]. While the kagome lattice materials have existed for a long time, $AV_3Sb_5$ ($A$ = K, Rb, and Cs) with a vanadium kagome lattice was recently discovered to be a superconductor, in which many interesting phenomena [7-10], such as giant anomalous Hall effect [11,12], unconventional charge density wave (CDW) [13-16], electronic nematicity [17-19], and pair density wave [20,21], have been observed. Owing to these exotic phenomena, the $AV_3Sb_5$ family superconductor attracts considerable attention and is being intensively studied.

As a new superconductor with remarkable physical properties, one of the most important issues is the superconducting (SC) gap symmetry because it is fundamental in clarifying the microscopic pairing mechanism and interplays between multiple electronic orders. Previous high-resolution angle-resolved photoemission spectroscopy (ARPES) studies on niobium-substituted $Cs(V_{0.93}Nb_{0.07})_3Sb_5$ and tantalum-substituted $Cs(V_{0.86}Ta_{0.14})_3Sb_5$ suggest an isotropic and nodeless SC gap in both cases [22]. Such a SC gap is also corroborant by the observations of exponentially temperature-dependent magnetic penetration depth [23,24]. However, certain V-shaped gaps, as well as residual Fermi-level states measured by scanning tunneling spectroscopy [20,25,26] in $CsV_3Sb_5$ seem to support a scenario of the anisotropic SC gap structure. Moreover, it has been reported that the anisotropy in the SC gap of $CsV_3Sb_5$ can be suppressed by chemical substitution or electron irradiation [27,28]. Therefore, to pin down the gap symmetry, the direct measurement of the SC gap structure in pristine $CsV_3Sb_5$ is highly desired.

In this study, we have used ultra-high resolution and low-temperature laser-based ARPES to directly measure the SC gap in the momentum space of kagome superconductor $CsV_3Sb_5$ (more details in Methods). We choose high-quality single crystals that have a relatively higher SC transition temperature $T_c \simeq 3$ K to ensure the accuracy of the SC gap measurements (see Supplementary Fig. S1). Figure 1(a) illustrates the crystal structure of $CsV_3Sb_5$, which belongs to the space group P6/mmm, with vanadium atoms forming a kagome lattice. As shown in the phase diagram in Fig. 1(c), in addition to the SC transition, $CsV_3Sb_5$ undergoes a CDW transition around $T_{CDW} \simeq 94$ K. The SC phase has two domes with applying physical or chemical pressure [29-31], implying an unconventional interplay between SC and CDW orders. Figure 1(b) presents the schematic Fermi surface (FS) of $CsV_3Sb_5$, which includes a circular pocket and a hexagonal pocket around the center of the Brillouin Zone (BZ), and a triangular pocket around the BZ corner. By performing careful ARPES measurements along these three FS sheets, we observe an isotropic SC gap structure on the circular and triangular FS sheets. Strikingly, we find a strongly anisotropic SC gap on the hexagonal FS with the gap maximum along the V-V bond direction, originating from the 3*d*-orbital electrons of the V kagome lattice. This direct observation of the FS-dependent anisotropic Cooper pairing lays a foundation to comprehensively understand the pairing symmetry and mechanism of kagome superconductors.

## Results

We first map out the FS by integrating ARPES intensity near the Fermi level ($E_F$), which is crucial for accurately locating the momentum position where the SC gap measures, especially for a multiband material system. As shown in Fig. 2(a), the experimentally measured FS fits well with the calculated FS contours [32,33]. Following the previous study [22,34], we label the FS sheets in this study as α, β, and δ, corresponding to the central circular, hexagonal pockets, and outer triangular pockets, respectively. We note that unlike in previous ARPES studies, the splitting of the α FS is absent here due to its pronounced $k_z$ dependence [35-38]. Figure 2(b) shows the energy-momentum ($E$-$k$) map along the ΓK direction taken by *s*- and *p*-polarized light, respectively. Consistent with the previous studies [34,36,39], the band of the β FS has a stronger intensity with *s*-polarization, while the band of the δ FS has a stronger intensity with *p*-polarization, which is helpful to distinguish the bands of the β and δ FS sheets.

We then study the SC gap in the momentum space of $CsV_3Sb_5$. In Figs. 2(c)-(e), we show the symmetrized energy distribution curves (EDCs) at the Fermi momentum ($k_F$) positions along with the α, β, and δ FS sheets. These EDCs are taken at $T$ = 2 K, below $T_c$ of approximately 3 K. For clarity, the $k_F$ positions are marked as the FS angle $\varphi$ defined in Fig. 2(a). We note that despite the limited detectable momentum range of the 5.8-eV laser, the measured area here spans over a 60-degree range of the FS angle, exceeding the minimum requirement for six-fold symmetry. As shown in Figs. 2(c) and 2(e), the symmetrized EDCs for the α and δ FS sheets exhibit similar shallow features near $E_F$, indicating relatively isotropic SC gaps. Remarkably, the symmetrized EDCs for the β FS, shown in Fig. 2(d), reveal a strongly FS-angle-dependent SC gap. The maximum SC gap is observed at the $k_F$ positions along the ΓM direction (FS angles of -31° and -85°), while the minimum SC gap is observed at the $k_F$ position along the ΓK direction (FS angle of -60°). The significant difference in the SC gap on the β FS along these directions is further illustrated by the EDCs without being symmetrized shown in Figs. 2(f) and 2(g). As shown in Fig. 2(f), along the ΓK direction, the leading edge of the EDC at $T = 2$ K shows almost no shift near $E_F$ compared to the EDC of a reference gold sample taken at the same condition, indicating a tiny gap or possible node. By fitting the EDC to a Bardeen-Cooper-Schrieffer (BCS) spectral function (Supplementary Note 2), we find the SC gap along the ΓK direction is approximately 0.1 meV. Conversely, a noticeably larger SC gap is observed from the EDCs along ΓM at $T = 2$ K, as shown in Fig. 2(g), and a BCS spectral function fit gives out a gap value of approximately 0.53 meV.

Quantitively, we extracted the SC gap magnitude by fitting all the EDCs along three FSs at $T = 2$K. The fitted EDCs are plotted in Figs. 2(c)-(e) in a symmetrized way, and the fitted SC gap magnitudes are summarized in Fig. 3(b) as a function of the FS angle. The corresponding $k_F$ positions are shown in Fig. 3(a) as open black circles. As shown in Fig. 3(b), the SC gap structure of the β FS has a large anisotropy, approximately 80%, while the SC gaps on the α and δ FS sheets are isotropic. Consistently, the similar anisotropic gap structure of the β FS is independently revealed by parameter-free analyses based on the leading-edge of the EDCs, as described in Supplementary Note 3 and Figs. S2-3. These results are further validated by their reproducibility, as the same SC gap structures are observed in independent samples with different alignments, with the details provided in Supplementary Figs. S4-7. Assuming the rotation symmetry remains intact [40], we symmetrize the SC gap distribution on the measured $k_F$ positions following the six-fold symmetry and present the SC gap structure of in-plane momentum space in Fig. 3(c). It is of notice that integrating the SC gaps across the three FS sheets maintains the anisotropy, as the isotropic SC gaps on the α and δ FS sheets do not introduce distinctive features. Therefore, our ARPES results align with the overall anisotropic SC gap structure deduced from the magnetic penetration depth measurements [27] and the V-shaped local density of states measured by scanning tunneling spectroscopy [20,25,26].

**Discussion**

We then discuss the pairing mechanism underlying the observed SC gap structure in $CsV_3Sb_5$. The anisotropic SC gap is selectively observed on the β FS, which is mainly occupied by 3*d* electrons of the V atom in the kagome lattice, suggesting possible unconventional pairing states [4,41-45]. Specifically, the pairing states mediated by bond-order fluctuations due to the geometrical frustration of the kagome lattice, including anisotropic (or nodal) *s*-wave or chiral *p*/*d*-wave [41] or even more exotic *f*-wave [42,46], which are characterized by an enhanced SC gap along the V-V bond direction (ΓM) and a gap minimum (or node) along the ΓK direction, matches well with the observed anisotropic SC gap on the β FS. The δ FS also contains V 3*d* orbital components but exhibits an isotropic SC gap, mainly because of the CDW gap [34,36,47,48]. The band-structure calculations [49] demonstrate the mixed occupation from V 3*d* and Sb 5*p* electrons on the β and δ FSs, and the anisotropic CDW gaps deplete the density of V 3*d* electrons most significantly on the δ FS while having fewer effects on the β FS (Supplementary Note 5 and Fig. S8). Consequently, the isotropic SC gap on the δ FS likely results from residual Sb 5*p* electron pairing, whereas the anisotropic gap on the β FS is predominately due to the V 3*d* electron pairing. While the V 3*d* electron pairing shows a strong anisotropy, the Sb 5*p* electron pairing, which causes isotropic SC gap on the α and δ FSs, seems consistent with the *s*-wave pairing driven by strong electron-phonon couplings reported in $CsV_3Sb_5$ [34,47]. However, we note that generally if superconductivity is realized at the same temperature for all FSs, their symmetries should all be the same. This implies that even if the β FS has gap nodes, these nodes cannot be symmetry-derived, such as *f*-wave, because they would be incompatible with isotropic gaps on the α and δ FSs. In this sense, anisotropic s-wave or chiral p/d-wave is more consistent with the observed FS-dependent SC gap anisotropy.

It is worth noting that, as shown in Fig. 1(c), the anisotropy in the V 3*d*-orbital SC gap can be significantly reduced by chemically substituting V with 7% Nb or 14% Ta [22]. This behavior contrasts with the robust gap anisotropy (*d*-wave) observed in cuprate superconductors, which remains unaffected by chemical substitutions [50-52]. The isovalent Nb/Ta substitutions neither introduce carriers nor change the lattice geometry, despite causing a minor expansion of the in-plane lattice (less than 1.2%) [22]. This rapid suppression of the SC gap anisotropy upon Nb/Ta substitutions suggests the gap anisotropy in $CsV_3Sb_5$ might not be imposed by the lattice symmetry. Instead, the suppression of the CDW upon the 14%-Ta upon 7%-Nb substitutions points to a crucial interplay between CDW and superconductivity in shaping the SC gap anisotropy. Particularly, the CDW in $CsV_3Sb_5$ exhibits a 2×2 in-plane charge density modulation [16,20,25,26], and below $T_c$, the emergence of pair density wave (PDW) was theoretically proposed [53-55] and a 2×2 PDW was experimentally demonstrated recently by scanning tunneling spectroscopy studies on $KV_3Sb_5$ and $CsV_3Sb_5$ [20,56].

Alternatively, the observed SC gap structure can be also explained by such a 2×2 PDW in $CsV_3Sb_5$. This order folds the Bogoliubov quasiparticle and introduces a PDW gap where the folded FS overlaps with the original FS. As the folded FS illustrated in Fig. 3(d), on the β FS, the intra-band overlaps occur primarily near the M point, leading to an enhanced SC gap in that region and contributing to the anisotropy in the SC gap. In contrast, there are no intra-band overlaps on the α and δ FS sheets, the SC gap remains unaffected and isotropic. Thus, the modulation of the pairing density induced by a 2×2 PDW contributes to the anisotropic SC gap structure selectively along the β FS, as shown in Fig. 3(c). Supporting this, a recent self-consistent theory of 2×2 PDW in the kagome lattice [55] predicts a highly similar gap anisotropy with that observed on the β FS. Furthermore, with chemical substitutions, the PDW can be rapidly suppressed, leading to a reduction in the anisotropy of the SC gap. While the nature of the 2×2 PDW in $CsV_3Sb_5$ is not fully understood and warrant further experimental and theoretical investigation, the observed SC gap structure in momentum space shows good consistency with the PDW in real space, providing additional insights into this phenomenon.

In summary, we have investigated the SC gap distribution in the momentum space of kagome superconductor $CsV_3Sb_5$ by high-resolution laser ARPES measurements. In addition to finding an isotropic SC gap on the FS derived from Sb 5*p* electrons, we identify a highly anisotropic SC gap with anisotropy over 80% occurring selectively on the FS occupied by 3*d* electrons of the V kagome lattice. This anisotropic gap reaches a maximum along the V-V bond direction (ΓM) and a minimum along the ΓK direction. Our direct observation of the FS-dependent anisotropic Cooper pairing in pristine $CsV_3Sb_5$ points to the unconventional pairing symmetries and could also be consistent with the SC gap structure of a 2×2 PDW, providing a foundation for further understanding the nature of kagome superconductivity. Combined with the previous observations of the vanished gap anisotropy upon modest Nb/Ta substitutions of V in the kagome lattice, our results open new prospects for the theorists because they would find which parameters drive the transition of isotropic-anisotropic superconductivity.

## Methods

### Growth of single crystals

High-quality single crystal $CsV_3Sb_5$ were synthesized from Cs bulk (Alfa Aesar, 99.8%), V pieces (Alddin, 99.97%), and Sb shot (Alfa Aesar, 99.999%) via a self-flux method using $Cs_{0.4}Sb_{0.6}$ as the flux. The above starting materials were put into an aluminum crucible and sealed in a quartz tube, which was then heated to 1000 °C in 24 hours and maintained for 200 hours. After that, the tube was cooled to 200 °C at a rate of 3 °C/hour, followed by cooling down to room temperature with the furnace switched off. To remove the flux, the obtained samples were soaked in deionized water. Finally, shiny single crystals with hexagonal features were obtained.

### High-resolution laser-ARPES measurements

Ultrahigh-resolution ARPES measurements were performed in a laser-based ARPES setup at ISSP, the University of Tokyo, which consisted of a continuous wave (CW) laser (hυ = 5.8 eV) provided from OXIDE Corporation and a vacuum ultra-violet quasi-CW laser (hυ = 6.994 eV), a Scienta HR8000 hemispherical analyzer and a sample manipulator cooled by decompression-evaporation of the liquid helium [57]. The CW laser with a photon energy of 5.8 eV was used to measure the superconducting gap, and the energy resolution was better than 0.6 meV. Its photon energy of 5.8 eV is the highest at which stable and sufficiently strong laser output can be obtained under CW conditions. The samples were cleaved *in-situ* and measured under a vacuum better than $3 \times 10^{-11}$ torr. The angular resolution was 0.1°. The Fermi level $E_F$ was calibrated with an *in-situ* connected gold reference.

**Data availability:** All relevant data are available from the corresponding author K. Okazaki (okazkai@issp.u-tokyo.ac.jp) upon reasonable request.

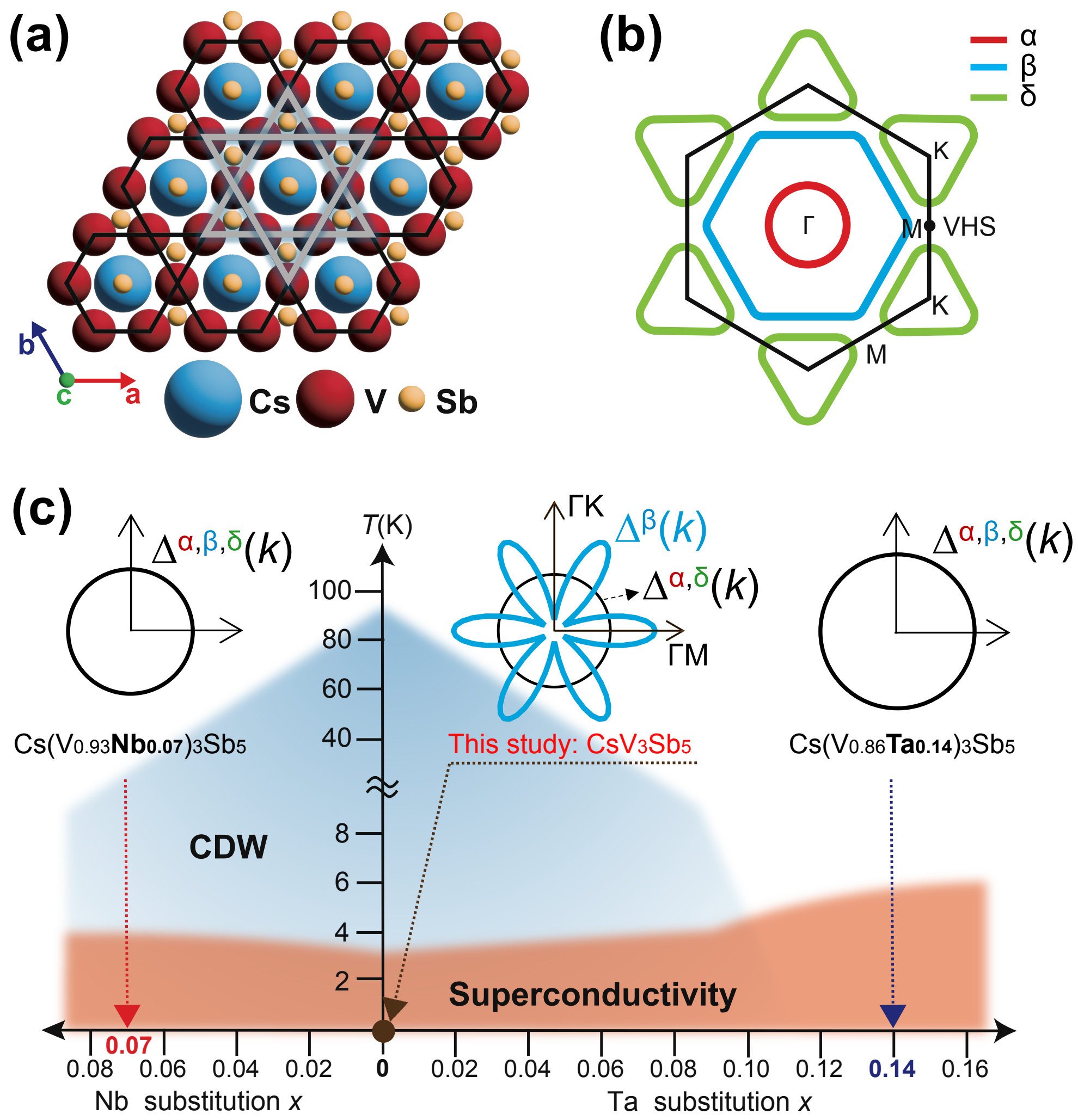

**Fig. 1. Crystal structure, Fermi surface topology, and superconducting phase diagram of $CsV_3Sb_5$. (a)** Crystal structure of $CsV_3Sb_5$ viewed from c-axis. The thick gray lines highlight the kagome lattice formed by V atoms. **(b)** Schematic FS contour of $CsV_3Sb_5$ and the van Hove singularities (VHS) at the M point is highlighted. **(c)** Phase diagram with Nb and Ta substitution of V atoms. Red and blue dashed arrows indicate the locations in the phase diagram for substituted samples $Cs(V_{0.93}Nb_{0.07})_3Sb_5$ and $Cs(V_{0.86}Ta_{0.14})_3Sb_5$, which exhibit the isotropic superconducting gap reported in the previous study by Zhong *et al* [22]. The anisotropic superconducting gap on the β FS and isotropic gap on the α and δ FS sheets in $CsV_3Sb_5$ revealed in this work are schematically plotted in the middle of the top panel in **(c)**.

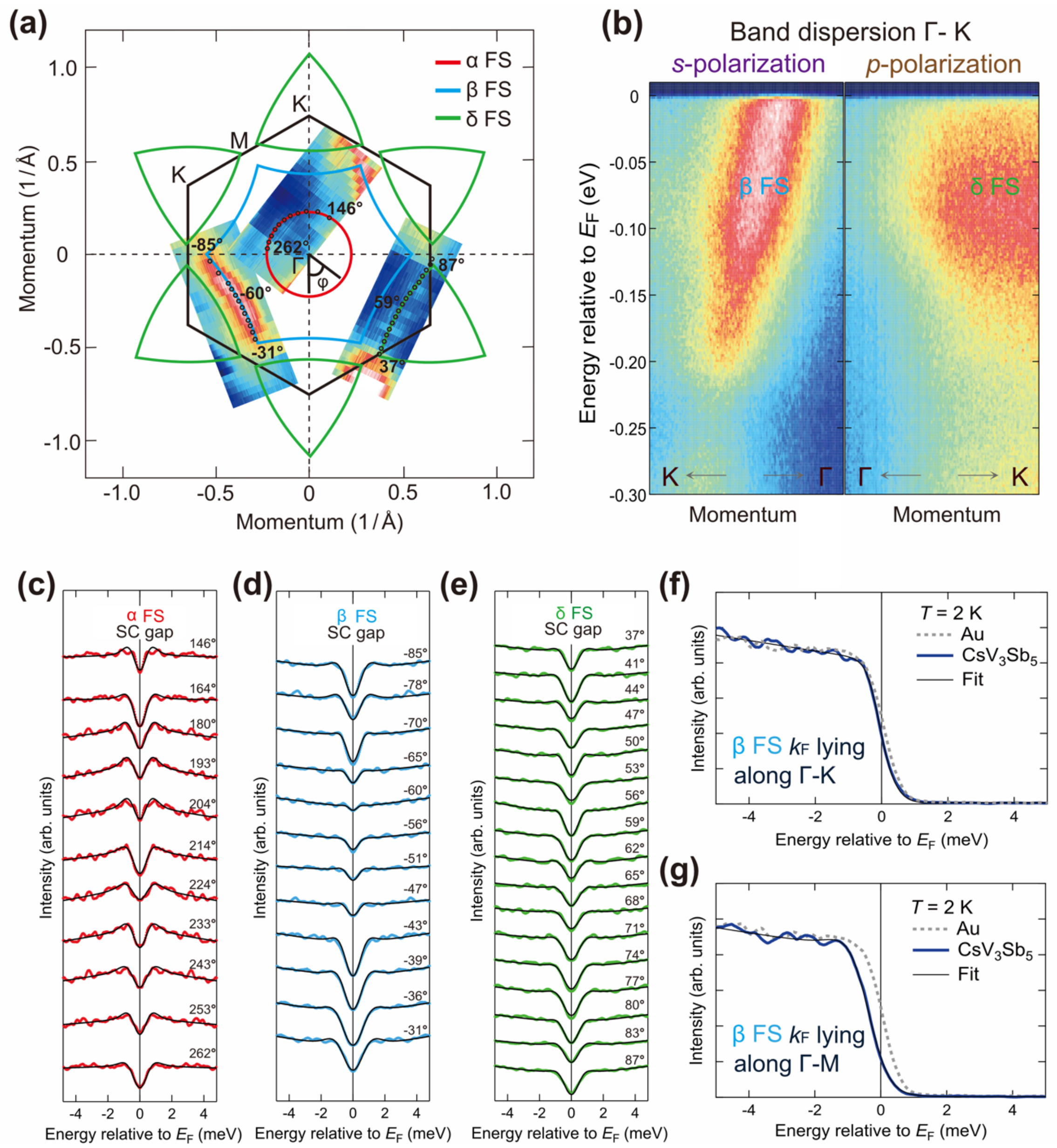

**Fig. 2. Fermi surface mapping and superconducting gap anisotropy. (a)** FS mapping of $CsV_3Sb_5$. Colored lines represent the FS contours from the band-structure calculations. The black open circles correspond to the $k_F$ positions where the EDCs are plotted. **(b)** Energy-momentum (*E*-*k*) maps along the ΓK direction obtained with *s*- and *p*-polarization, respectively. **(c-e)** Symmetrized EDCs measured at $T = 2$ K in different $k_F$ positions for the α, β, and δ FS sheets, respectively. The $k_F$ positions are marked as FS angle ($\varphi$) as defined in **(a)**. **(f), (g)** EDCs at $k_F$ points lying along the ΓK and ΓM directions on the β FS, respectively. Black lines on the top of the EDCs at 2 K are the best fits to the BCS spectral function. To better visualize the gap opening, the EDC of the reference sample gold are plotted as the dashed gray lines.

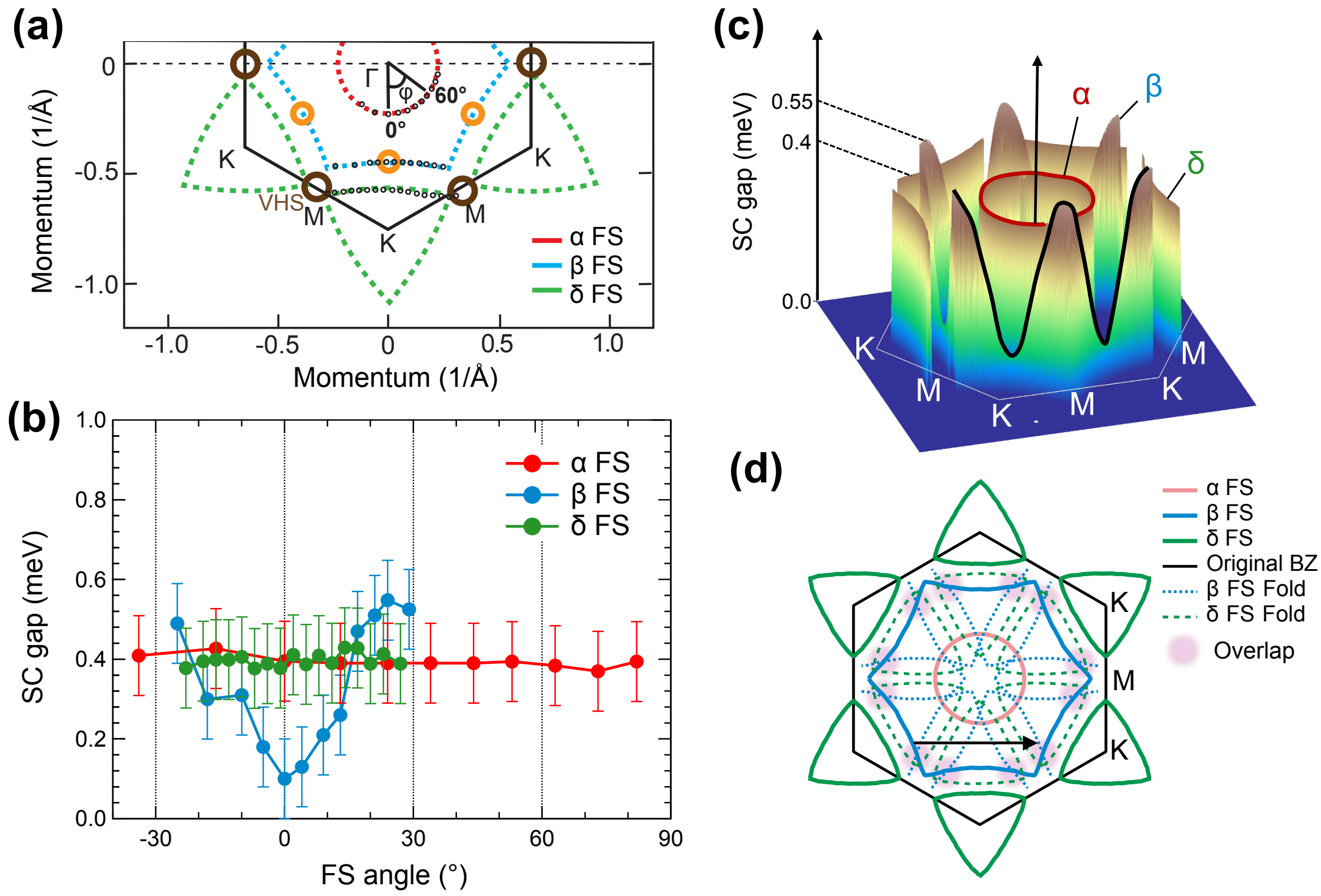

**Fig. 3. Superconducting gap structure and pairing modulation induced by 2×2 pairing density wave. (a)** A summary of the $k_F$ positions, where the SC gap is measured. The dashed color lines represent the FS contours of $CsV_3Sb_5$. For better visualization, the $k_F$ points were rotated using the system's six-fold symmetry to align them into a single sector. The yellow circles mark the momentum positions with a minimum SC gap, while the brown circles mark the momentum location of van Hove singularities (VHS). **(b)** Fitted SC gap magnitudes on three FS sheets plotted as a function of FS angle ($\varphi$) as defined in **(a)**. The uncertainties in the gap values were determined by the stability of the Fermi level, which is evaluated to be ±0.1 meV in our ARPES setup. **(c)** Schematic plot of the SC gap distribution in the in-plane momentum space of $CsV_3Sb_5$. **(d)** Folded Bogoliubov quasiparticle states resulting from the 2×2 pairing density wave. The pinked shades highlight the intra-band overlaps between the folded and original β FS.